\documentclass[authoryear,preprint,10pt,review]{elsarticle}
\usepackage{amssymb,amsmath,gensymb,multirow, threeparttable}
\usepackage{lineno}

\makeatletter
\def\bbordermatrix#1{\begingroup \m@th
  \@tempdima 4.75\p@
  \setbox\z@\vbox{%
    \def\cr{\crcr\noalign{\kern2\p@\global\let\cr\endline}}%
    \ialign{$##$\hfil\kern2\p@\kern\@tempdima&\thinspace\hfil$##$\hfil
      &&\quad\hfil$##$\hfil\crcr
      \omit\strut\hfil\crcr\noalign{\kern-\baselineskip}%
      #1\crcr\omit\strut\cr}}%
  \setbox\tw@\vbox{\unvcopy\z@\global\setbox\@ne\lastbox}%
  \setbox\tw@\hbox{\unhbox\@ne\unskip\global\setbox\@ne\lastbox}%
  \setbox\tw@\hbox{$\kern\wd\@ne\kern-\@tempdima\left[\kern-\wd\@ne
    \global\setbox\@ne\vbox{\box\@ne\kern2\p@}%
    \vcenter{\kern-\ht\@ne\unvbox\z@\kern-\baselineskip}\,\right]$}%
  \null\;\vbox{\kern\ht\@ne\box\tw@}\endgroup}
\makeatother

\journal{: arXiv} 
\begin{document}
\begin{frontmatter}
\title{Revised calculation of the coefficient of parentage in plant breeding}

\author[UFG]{Carlos Hernandez-Suarez\corref{cor1}}
 \ead{cmh1@cornell.edu}
 
 \author[UC]{Osval Montesinos L\'opez\corref{cor2}}
 \ead{oamontes1@ucol.mx}

\address[UFG]{Instituto de Innovaci\'on y Desarrollo, Universidad Francisco Gavidia, San Salvador, El Salvador}
\address[UC]{Facultad de Telem\'atica, Universidad de Colima, Colima 28040, Mexico}
  
 \cortext[cor1]{Corresponding author}

\begin{abstract}
The Coefficient of Parentage (COP) between two individuals is the expected inbreeding of their offspring. Originally exploited by animal breeders, is now a routine calculation among plant breeders as part of crop improvement programs. Here we show that the COP between strains requires a different calculation than the used to calculate the COP between individuals. Failure to do so may result in an overestimation of the amount of inbreeding. Here we provide a simple methodology to calculate the correct coefficient of parentage between strains.

\end{abstract}


\end{frontmatter}


\section{Introduction}

The Coefficient of Parentage (COP) between two individuals \textit{X} and \textit{Y}  is the probability that an allele taken from a particular locus in individual \textit{X} is identical by descent to an allele taken from the same locus in individual \textit{Y}. It was originally defined by  \cite{wright1922coefficients} who developed it in terms of correlation analysis, whereas \cite{malecot1948} used a probabilistic approach. Twice the COP value determines the additive genetic correlation between two strains, which is known as the \textit{Coefficient of Relationship} and it is a well-known estimate for analyzing breeding strategies. COP values computed from pedigree records have been used extensively in livestock improvement since the 70's \cite{mrode1996linear}. The COP is in fact a measure of relatedness in terms of some reference, parent population. A typical reference population would be one without previous pedigrees at the time of foundation, where it is assumed that all founders are unrelated and not inbred.

Although several indices have been proposed to estimate genetic diversity based on different kind of data ranging from morphological to molecular traits, the classic COP is currently used by not only by animal breeders but also plant breeders as a prerequisite for any cultivar development program. Several studies have shown the usefulness and power of COP values for crop improvement; for instance, \cite{cox1985relationship} demonstrated the correlation between COP values and similarity indices based on isozyme markers in Soybean (\textit{Glycine max}, (L) Merr.) \cite{murphy1986cluster} studied the population structure and field diversity of red winter wheat (\textit{Triticum aestivum} L.) cultivars based on COP values. Several cultivar improvement programs of important crops such as wheat, rice and sugar have been based on pedigree analysis and COP values \cite{souza1994spring, dilday1990contribution, lima2002analysis}.

More recently, \cite{bernardo2002breeding} demonstrated the utility of COP values for predicting single cross performance in hybrid breeding programs of Maize.  Additionally, more direct applications of COP values for modeling additive and additive $\times$ additive components of genetic variance as well as concomitant components of genotype by environmental variation in crop breeding evaluation data are now being proposed \cite{crossa2006modeling, burgueno2007modeling}.

In plant breeding, a strain is a sample of seeds, either germplasm from a genebank or other collection, or seeds collected from a breeding process. It is not synonymous with genotype since it may be a mixture of individuals with different genotypes. The COP between two strains of germplasm \textit{X} and \textit{Y} is the probability that a randomly selected allele from one strain is identical by descent to a randomly selected allele at the same locus from the other strain \cite{falconer1996introduction}. By ``randomly selected allele" it is meant observing the allele at a particular locus from a randomly selected individual in the sample, that is, sampling of an individual's allele from the sample of individuals.

This definition, as we will show, requires a different calculation of the COP when subjects (nodes) in a pedigree are either individuals or strains, because in the first case we deal with allele probabilities and in the second with allele frequencies. In other words, here we state that in plant breeding the average inbreeding of the $F_1$ of \textit{X} and \textit{Y} is not always reflected by the COP between those strains, when the COP is calculated according to the methods designed in animal breeding. 

In this paper we will denote $f _{XY}$ for the COP between two individuals and $f^*_{XY}$ to the COP between two strains.

\section{Materials and Methods}

\subsection{Review of calculation of COP for individuals}

\begin{figure}[htbp]
\begin{center}
\includegraphics[width=12cm]{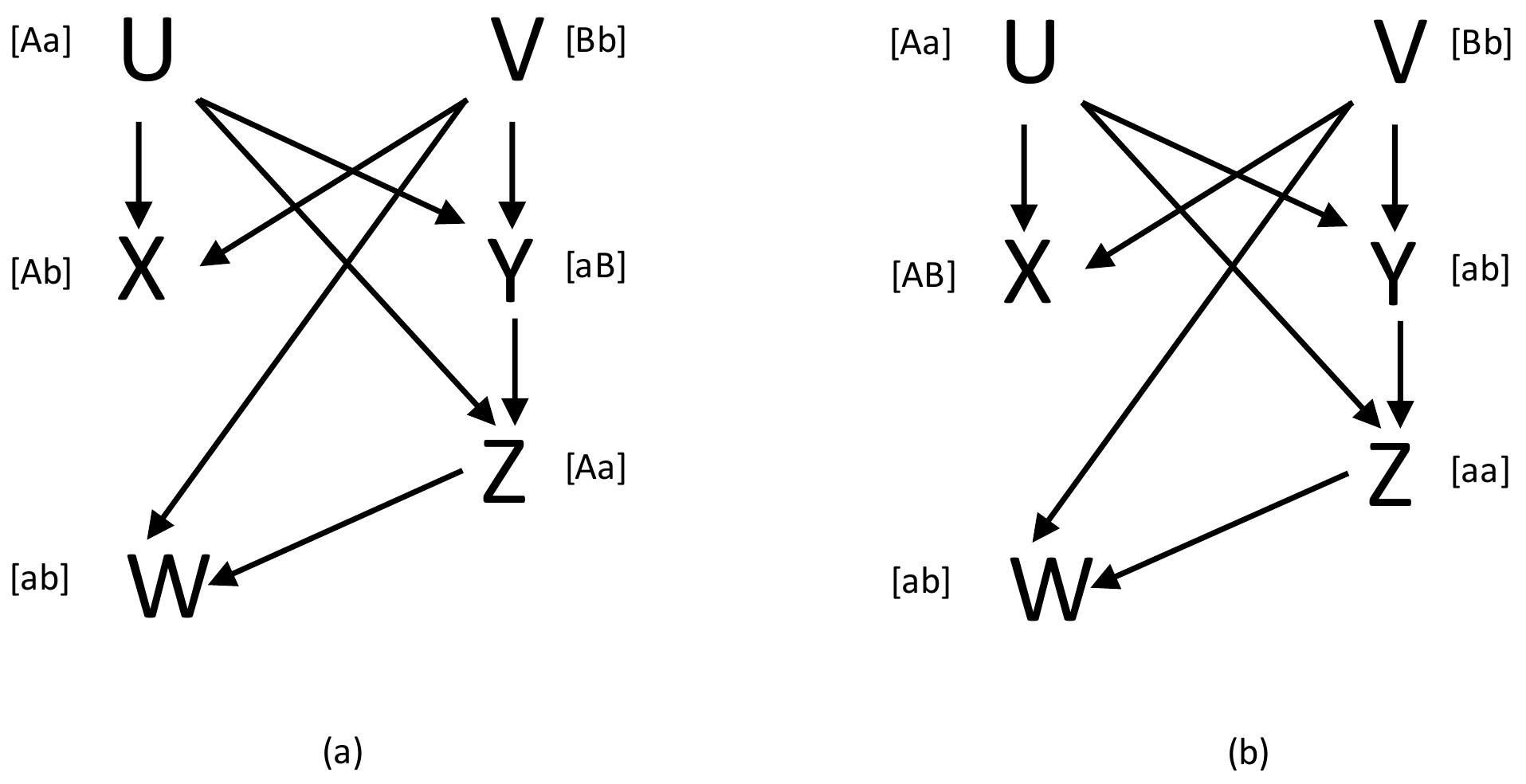}
\end{center}
\caption{Two realizations of allele inheritance in a particular pedigree. In (a) the probability that two alleles taken from the same locus from individuals \textit{Y} and \textit{W} are identical by descent is $1/4$, whereas in (b) it is $1/2$. The COP between \textit{Y} and \textit{W} is calculated considering all possible genotypes for \textit{Y} and \textit{W}.\label{fig1}}
\end{figure}

Fig.~\ref{fig1} shows two simulations of allele inheritance through the nodes of a pedigree, where the nodes are individuals. In Fig.~\ref{fig1}a the probability that two alleles taken from the same locus from individuals \textit{Y} and \textit{W} are identical by descent is $1/4$, whereas in Fig.~\ref{fig1}b it is $1/2$. The COP between \textit{Y} and \textit{W} should be calculated by including all possibilities of allele inheritance for these two individuals. An important assumption is that there is no selection for a particular allele or set of alleles, that is, each node inherits a particular allele to a descendant with probability $1/2$.

\begin{figure}[htbp]
\begin{center}
\includegraphics[width=12cm]{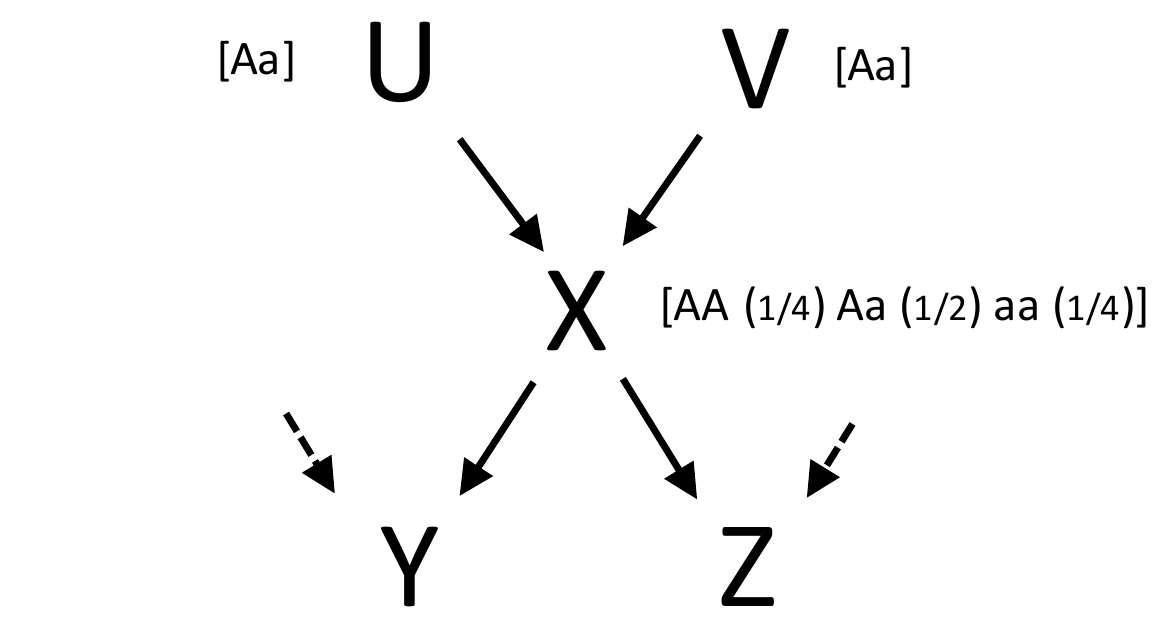}
\end{center}
\caption{An example pedigree. The possible genotypes for individual \textit{X} together with their probabilities are shown. A parent of each of \textit{Y} and \textit{Z} is not shown and they are assumed to be unrelated to any other individual in the pedigree and thus they do not contribute to $f_{YZ}$.\label{fig2}}
\end{figure}

In Fig.~\ref{fig2} we can see a pedigree involving five individuals.  Next to each individual there is the possible genotypes together with the probabilities for each genotype. A parent of each of \textit{Y} and \textit{Z} is not shown and assumed to be unrelated to the rest of the individuals of the pedigree and do not contribute to the COP between \textit{Y} and \textit{Z}. Their alleles are denoted by ``-''. Table~\ref{table1} reviews the basic calculation behind the COP between individuals \textit{Y} and \textit{Z}.

\begin{table}[ht]
\centering
\caption{{\bf Calculation of the COP for pedigree in Fig.~\ref{fig2}}. $P_1=$ probability of the specific allele combination, $P_2=$ probability that two alleles one of each of \textit{Y} and \textit{Z} will be identical, $P_3 =P_2 \times P_3$. The COP between \textit{Y} and \textit{Z} is the sum of column $P_3$ equal to $3/16$. The symbol ``-''' indicates an allele that plays no role in the COP between \textit{Y} and \textit{Z}.}
\label{table1}
\begin{tabular}{|c|c|c|c|c|}
\hline
\textit{Y}   & \textit{Z}   & $P_1$ & $P_2$ & $P_3$  \\ \hline
A - & A - & $3/8$ & $1/4$ & $3/32$ \\ \hline
A - & a - & $1/8$ & $0$   & $0$    \\ \hline
a - & A - & $1/8$ & $0$   & $0$    \\ \hline
a - & a - & $3/8$ & $1/4$ & $3/32$ \\ \hline
\end{tabular}
\end{table}

The inbreeding coefficient of an individual \textit{X}, namely $F_X$, is the frequency of homozygous loci. It can also be seen as the probability that a random chosen locus is homozygous. The use of $F_X$ in the calculation of the COP can be illustrated by generating an analogous table to Table~\ref{table1} as a function of $F_X$. The result is Table~\ref{table2}, where we can see that if $F_X$ is known, the COP between \textit{Y} and \textit{Z} is $(1+F_X)/8$. The use of the inbreeding coefficient simplifies the calculations because if this is known for the most immediate common ancestors of two individuals one can derive expressions in a similar way to those derived for Table~\ref{table2}, without having to look the pedigree up to the terminal ancestors. 

\begin{table}[ht]
\centering
\caption{{\bf Calculation of the COP for pedigree in in Fig.~\ref{fig2}} using the inbreeding coefficient of individual \textit{X}. The COP between \textit{Y} and \textit{Z} is $(1+F_X)/8$. Since $F_X=1/2$, the COP is $3/16$.}
\label{table2}
\begin{tabular}{|c|c|c|c|c|}
\hline
Y   & Z   & $P_1$        & $P_2$ & $P_3$  \\ \hline
A - & A - & $(1+F_X)/4$  & $1/4$ & $(1+F_X)/16$ \\ \hline
A - & a - & $(1-F_X)/4$      & $0$   & $0$    \\ \hline
a - & A - & $(1-F_X)/4$      & $0$   & $0$    \\ \hline
a - & a - & $(1+F_X)/4$ & $1/4$ & $(1+F_X)/16$ \\ \hline
\end{tabular}
\end{table}

\subsection{COP for strains}

While in an animal pedigree \textit{X} is an individual, in plant breeding \textit{X} is a strain, that is, a population composed by individuals that may exhibit genetic variability at a randomly chosen locus. The COP in this case must consider the probability that two alleles, taken each from the same locus from two \textit{randomly chosen individuals} from populations \textit{Y} and \textit{Z} are identical by descent. This extra sampling step adds some independence that will be explored later. In other words, the numbers between brackets next to the individual \textit{X} in Fig.~\ref{fig2} that correspond to the probability of each genotype, become genotype frequencies when \textit{X} is a population. This implies $F_X$ can not be used to reveal the probability that \textit{Y} and \textit{Z} share common alleles for the following reason: an $F_X$ value of $1$ implies that individual \textit{X} will pass only one type of allele to the descendants, whereas if \textit{X} is a population, both genotypes \textit{AA} and \textit{aa} exists in \textit{X} and both alleles \textit{A} and \textit{a} will be inherited to the descendant population. 

\begin{figure}[htbp]
\begin{center}
\includegraphics[width=10cm]{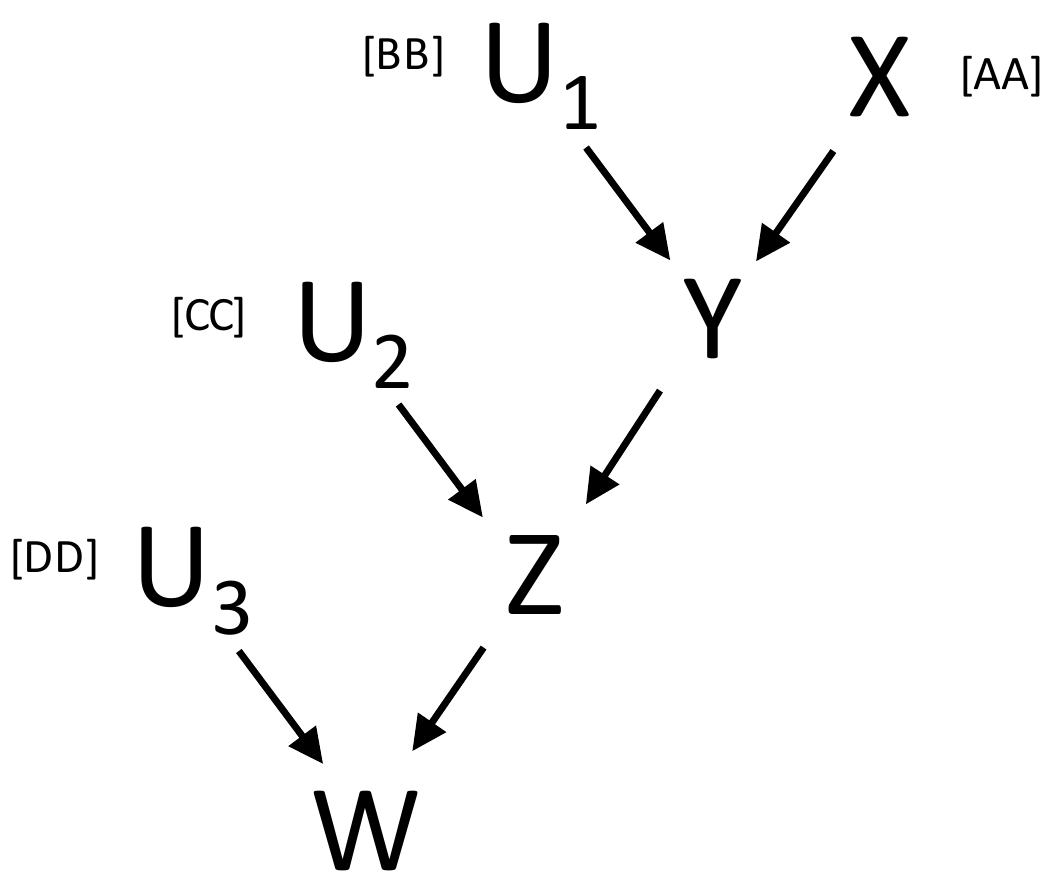}
\end{center}
\caption{Basic pedigree to exhibit differences in COP when nodes are individuals or strains. Original ancestors are $U_1,U_2,U_3$ and $X$ which are assumed to be homozygous.\label{fig3}}
\end{figure}

Pedigree in Fig.~\ref{fig3} will allow us to exhibit the differences in the COP between two nodes in a particular pedigree, depending these represent individuals or populations. Tables \ref{table3} and \ref{table4} show the allele composition for the cases nodes of Fig.~\ref{fig3} being individuals or populations respectively. 

\begin{table}[ht]
\centering
\caption{{\bf Genotype composition when nodes in pedigree of Fig.~\ref{fig3} are individuals}. Numbers in brackets are genotype probabilities.}
\label{table3}
\begin{tabular}{|c|c|l|l|l|}
\hline
Node & \multicolumn{4}{c|}{Genotype [Conditional probability]}                                               \\ \hline
Y    & \multicolumn{4}{c|}{AB {[}1{]}}                                                      \\ \hline
Z    & \multicolumn{2}{c|}{AC {[}1/2{]}}                & \multicolumn{2}{c|}{BC {[}1/2{]}} \\ \hline
W    & \multicolumn{1}{l|}{AD {[}1/2{]}} & CD {[}1/2{]} & BD {[}1/2{]}    & CD {[}1/2{]}    \\ \hline
\end{tabular}
\end{table}

\begin{table}[ht]
\centering
\caption{{\bf Genotype composition when nodes in pedigree of Fig.~\ref{fig3} are populations}. Numbers in brackets are genotype frequencies.}
\label{table4}
\begin{tabular}{|c|c|l|l|l|}
\hline
Node & \multicolumn{4}{c|}{Genotype of population [frequencies]}                       \\ \hline
Y    & \multicolumn{4}{c|}{AB {[}1{]}}                               \\ \hline
Z    & \multicolumn{4}{c|}{AC {[}1/2{]}, BC {[}1/2{]}}               \\ \hline
W    & \multicolumn{4}{l|}{AD {[}1/4{]}, CD {[}1/2{]}, BD {[}1/4{]}} \\ \hline
\end{tabular}
\end{table}

In Table~\ref{table3} it can be seen for instance that  individual \textit{Z} may be of genotype $AC$ with probability $1/2$ in which case individual \textit{W} will have genotypes $AD$ or $CD$ equally likely. The COP between individuals \textit{Z} and \textit{W} is clearly $1/4$ which is the sum of the probabilities that both alleles are $A \ (1/16)$, $B \ (1/16)$ or $C \ (1/8)$. 

In contrast, in Table~\ref{table4} it can be seen for instance that  population \textit{Z} has genotypes $AC$ and $BC$ at equal frequencies. From here, the allele frequencies of  \textit{A}, $B$ and $C$ are respectively $1/4,1/4$ and $1/2$ respectively. Population \textit{W} has alleles $A,B,C$ and $D$ at frequencies $1/8,1/8,1/4$ and $1/2$ respectively. Taking one allele from \textit{Z} and one from \textit{W} will yield the same allele with probability $3/16$, which is the COP between \textit{Z} and \textit{W}. Thus, we conclude that for the same pedigree of Fig.~\ref{fig3}, $f_{ZW}= 1/4$ whereas $f^*_{ZW}=3/16$. Notice that we observed the definition of COP in both cases.

\subsection{COP under selection}

We already showed that while the level of inbreeding is relevant to calculate the COP between individuals, it is irrelevant for populations. This is important in plant breeding, where the search for inbreeds is a common practice. Nevertheless, applying selection pressure is a different matter. The calculations involving the COP as originally conceived were never designed to consider selection pressure because it changes the basic assumption that an allele in an individual will be inherited to a descendant with probability $1/2$. Depending on the type of selection, one could calculate the COP according to its definition, for instance, if terminal ancestors are fully inbred and selection pressure is applied to one of the alleles at every cross, then at the end, the populations in the pedigree will be isogenic with homozygous individuals, and, at every loci, individuals are identical to the same loci of one of the common ancestors. We do not to attempt to explore far beyond this point, suffices to notice that including selection would require different calculations than those originally designed for individuals, or the introduced here, for populations with no selection.

\subsection{An additional example comparison}

\begin{figure}[htbp]
\begin{center}
\includegraphics[width=8cm]{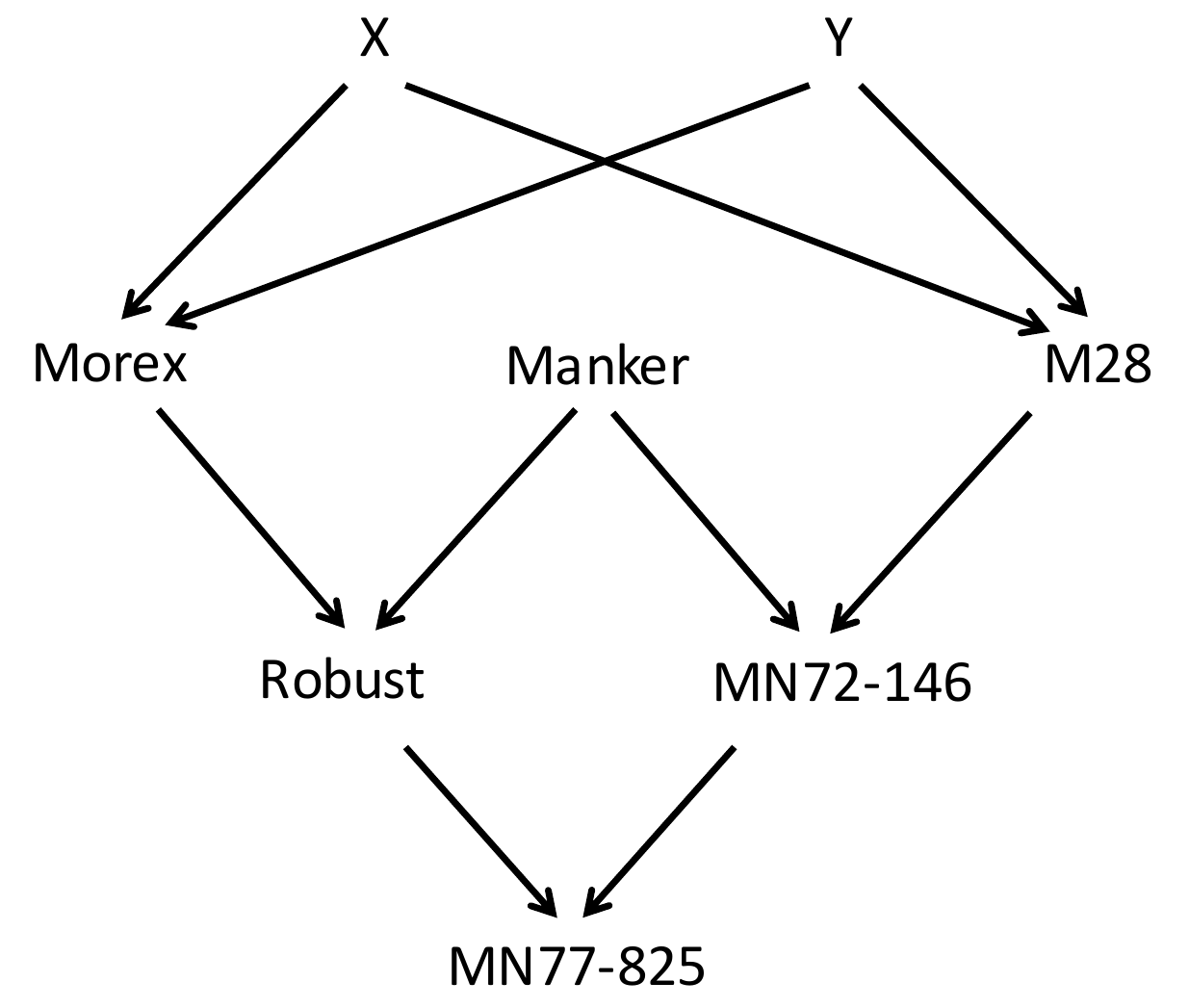}
\end{center}
\caption{Minnesota cultivars of malting barley. The pedigree has been modified from \cite[p. 44]{bernardo2002breeding}, by including strains \textit{X} and \textit{Y} to exhibit the fact that \textit{Morex} and \textit{M28} are full sibs. Strains \textit{X}, \textit{Y} and \textit{Manker} are assumed to be unrelated and fully inbred.\label{fig4}}
\end{figure}

Fig.~\ref{fig4} shows a pedigree with six cultivars of malting barley taken from \cite[p. 44]{bernardo2002breeding}. In this pedigree, \textit{Morex} and \textit{M28} are full sibs developed from different $F_2$ plants.

With the purpose of illustration we have included in this pedigree original ancestors \textit{X}  and \textit{Y} to exhibit the fact that \textit{Morex} and \textit{M28} are full sibs, which are assumed to be unrelated and inbred \cite[p. 44]{bernardo2002breeding}. with respective genotypes \textit{AA}  and \textit{BB}. The traditional calculation of the COP between strains in the pedigree of Fig. \ref{fig4} is here reproduced from \cite[p. 44]{bernardo2002breeding} yielding:

\begin{itemize}
\item[C1.]  The \textit{COP} between \textit{Morex} and \textit{Robust} comprise a parent-offspring relationship thus
$$
f_{(Morex,Robust)} = \frac{1}{2}[1+f_{(Morex,Manker)}] =  \frac{1}{2}
$$

\item[C2.]   \textit{Morex} and \textit{M28} are full-sibs thus the \textit{COP} between them, assuming their parents are unrelated is:
$$
f_{(Morex,M28)} = \frac{1}{2}[1+f_{(X,Y)}] =  \frac{1}{2}
$$

\item[C3.]   \textit{MN72-146} was developed from a cross between \textit{Manker} and \textit{M28}. Since these two are unrelated, the \textit{COP} between \textit{MN72-146} and either parent is:

$$
f_{(MN72-146,Manker)} = \frac{1}{2}[1+f_{(Manker,M28)}] =  \frac{1}{2}
$$

\item[C4.]   \textit{Robust} and \textit{MN72-146} are half sibs with \textit{Manker} as the common parent. The \textit{COP} between \textit{Robust} and \textit{MN72-146} is:

\begin{eqnarray*}
f_{(Robust,MN72-146)} &=& \frac{1}{4} \big[\frac{1}{2}[1+F_{Manker}]+f_{(Morex,Manker)}+\\
&  & f_{(Morex,M28)}+f_{(Manker,M28)}\big] \\
&=&\frac{1}{4}(1+0+\frac{1}{2}+0) =\frac{3}{8}
\end{eqnarray*}

\item[C5.]   \textit{Robust} and \textit{MN77-825} comprise a parent-offspring relationship, thus the \textit{COP} between \textit{Robust} and \textit{MN77-825} is:

$$
f_{(Robust,MN77-825)} = \frac{1}{2}[1+f_{(Robust, MN72-146)}] =  \frac{1}{2}[1+\frac{3}{8}]=\frac{11}{16}
$$

\end{itemize}

\subsection{Correct calculation of the inbreeding coefficient between strains for pedigree in Fig.~\ref{fig4}}

We can actually calculate the true value of the COP for each pair of strains according to the definition using a robust approach: since by assumption the genotype of \textit{X} is $[AA]$, that of \textit{Y} is $[BB]$  and \textit{Manker} is $[CC]$ (see Fig.~\ref{fig4}), then the genotype of \textit{Morex} is [$AB]$ and that of \textit{Robust} is $[AC,BC]$. The genotypes of the (\textit{Morex} $\times$ \textit{Robust})$F_1$ are $[AA,AC,AB,AC,BA,BC,BB,BC$] and only a quarter of these are homozygous, thus, the inbreeding coefficient of the offspring resulting from the cross of \textit{Robust} with its parent \textit{Morex} is $1/4$. 

We can achieve the same result by observing that the probability that an allele taken from \textit{Morex} (that is, from the set $[AB]$) and another from \textit{Manker } (that is, from the set $[AC,BC]$) are identical is $1/4$.  Nevertheless, in $C1$ above, it was found that the COP between \textit{Morex} and \textit{Robust} is $1/2$. Even if it is argued that \textit{Morex} and \textit{Robust} were selfed to become pure inbred and the COP between them was calculated after selfing, their genotypes would then be $[AA,BB]$ and $[AA,CC,CC,BB]$ respectively and the coefficient of inbreeding of the (\textit{Morex} $\times$ \textit{Robust}) $F_1$  is still $1/4$. 

The error is due, as mentioned before, to the use of expressions derived for pedigreed individuals: it is known that when \textit{Z} is the offspring of two unrelated individuals \textit{X} and \textit{Y} and \textit{X} has an inbreeding coefficient $F_X$, the COP between \textit{X} and \textit{Z} is known to be:

\begin{equation*}
f_{XZ} = \frac{1}{4}(1+F_X),
\end{equation*}
see Falconer (\citeyear{falconer1996introduction}, eq. 5.5) and Bernardo (\citeyear{bernardo2002breeding}, eq. 2.14). While this is true when \textit{X} and \textit{Z} are individuals, the expression has no place when \textit{X} and \textit{Z} are strains, since the degree of inbreeding of \textit{X} is irrelevant, what matters is the allele frequency in \textit{X} and \textit{Z}. In other words, an individual that is homozygous at a locus can only inherit one type of allele to the offspring; nevertheless, a strain may be homozygous and still can inherit several alleles to the offspring. This is the case of a strain that contains alleles $[AB]$ that is selfed several times to become fully inbred and thus containing individuals with genotypes \textit{aa} and $BB$. By definition, the strain is homozygous but inherits two alleles to the offspring. This is precisely the source of the mistake committed in $C1$ where it was assumed $F_X = 1$ because each strain was selfed to be considered inbred, yielding $f_{(Morex,Robust)} =  \frac{1}{2}$. 

Table~\ref{table1} shows the expected allele frequency for each strain in pedigree of Fig.~\ref{fig4}. Using data from this Table, the coefficient of inbreeding between strains can be obtained with the cross product of their relative allele frequencies. It can be verified that $f_{(Morex,Robust)} = \frac{1}{2}\frac{1}{4} +\frac{1}{2}\frac{1}{4} =  \frac{1}{4}$. Another example is the COP between \textit{Robust} and \textit{MN77-825}. According to $C5$ above, it is $11/16$, nevertheless, using Table~\ref{table1} it can be seen that the (\textit{Robust }$\times$ \textit{MN77-825)}$F_1$ has an inbreeding coefficient of $(1/4)^2+(1/4)^2+(1/2)^2=3/8$, regardless of the level of inbreeding of \textit{Robust} and \textit{MN77-825}.

\begin{table}[!ht]
\caption{{\bf Relative expected allele frequencies of each strain for pedigree in Fig.~\ref{fig4}.}}
\centering
\begin{tabular}{lclclclc|}
\\
\hline
Strain & \multicolumn{3}{c}{Allele}                                                                    \\ \hline
        & A   & B   & C   \\
        \hline
X        & 1   & 0   & 0   \\
Y        & 0   & 1   & 0   \\
Manker   & 0   & 0   & 1   \\
Morex    & 1/2 & 1/2 & 0   \\
M28      & 1/2 & 1/2 & 0   \\
Robust   & 1/4 & 1/4 & 1/2 \\
MN72-146 & 1/4 & 1/4 & 1/2 \\
MN77-825 & 1/4 & 1/4 &   1/2 \\
 \hline\end{tabular}
\label{table5}
\end{table}

\section{Results}

\begin{figure}[htbp]
\begin{center}
\includegraphics[width=9cm]{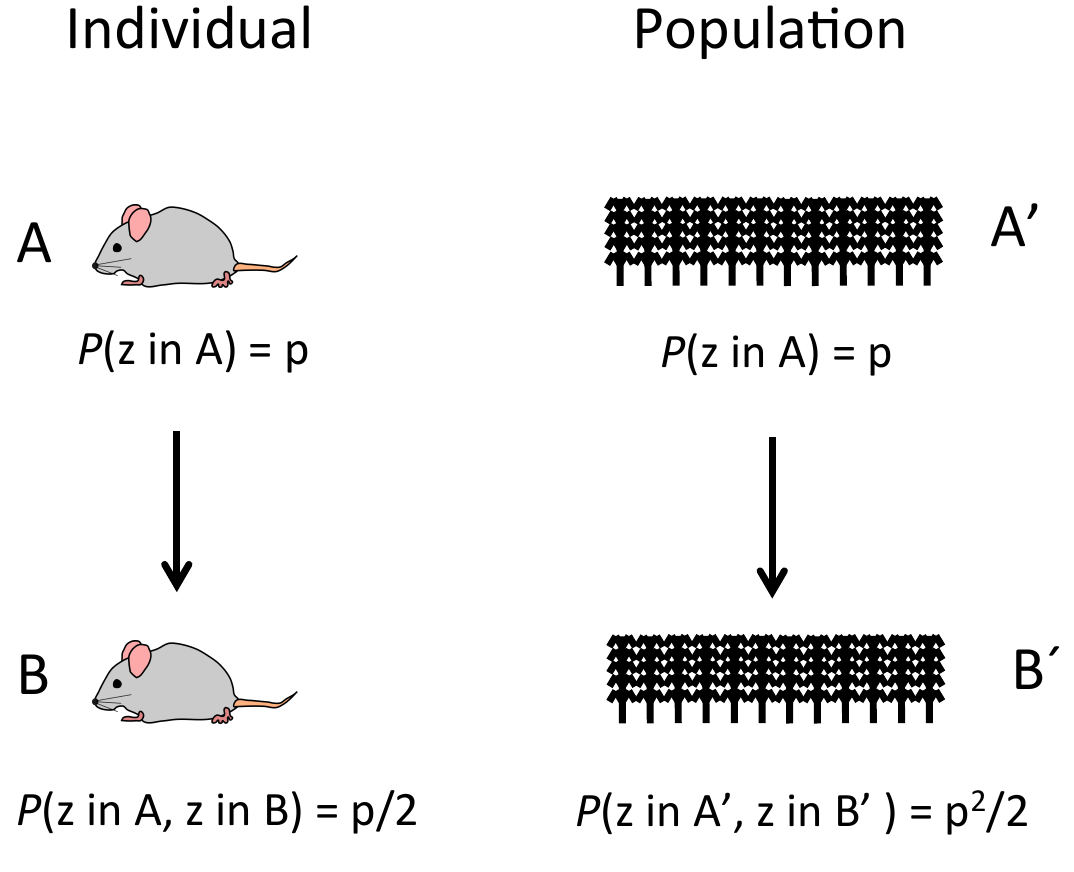}
\end{center}
\caption{Diagram showing the source of the difference in the calculation of the \textit{COP} between individuals or strains. In the left side we can see that if the probability that an individual $A$ has exactly one allele $z$ is $p$, it may inherit this allele to a descendant $B$ with probability $1/2$, and the probability that both contain allele $z$ is $p/2$. In the right side, every individual of the strain $A$ has a probability $p$ of having exactly one allele $z$, thus the relative frequency of the allele in strain $A$ is $p$ and that of descendant strain $B$ is $p/2$. The probability that two individuals $A'$ and $B'$ taken at random from each strain $A$ and $B$ will both have allele $z$ is $p^2/2$. Here we assume that parents of $B$ and $B'$ (not shown) do not have allele $z$.\label{fig5}}
\end{figure}

The reason why using formulae designed to be applied on pedigreed individuals may not yield the correct COP in strains has to do with the degree of independence between two particular events. Fig.~\ref{fig5} illustrates the difference when calculating COP between individuals or populations under no selection pressure: if the probability that \textit{A} has allele \textit{z} is \textit{p}, then the probability that \textit{B} has allele \textit{z} if \textit{A} has it is $1/2$, thus the probability that both \textit{A} and \textit{B} have allele \textit{z} is just $p/2$. Nevertheless, if \textit{A} and \textit{B} are strains, the probability that an individual in strain \textit{A} has allele \textit{z} is \textit{p} which is the expected frequency of allele \textit{z} in \textit{A}, whereas that of \textit{B} is $p/2$, thus, the probability that we take an individual at random from strain \textit{A} and another from \textit{B} and both have allele \textit{z} is $p^2/2$ (see Fig.~\ref{fig5}).

It is possible to show that the COP between strains is overestimated when calculated using formulae designed to work for individuals. Appendix A1 provides a mathematical proof that $f_{XY} \ge f^*_{XY}$.

\subsection{A simple algorithm to calculate the COP between strains in the absence of selection}

With the previous considerations in mind, we now introduce a method to calculate the expected COP when the nodes in a pedigree are strains. This method requires only that the allelic frequencies are not altered due to selection.

\begin{figure}[htbp]
\begin{center}
\includegraphics[width=6cm]{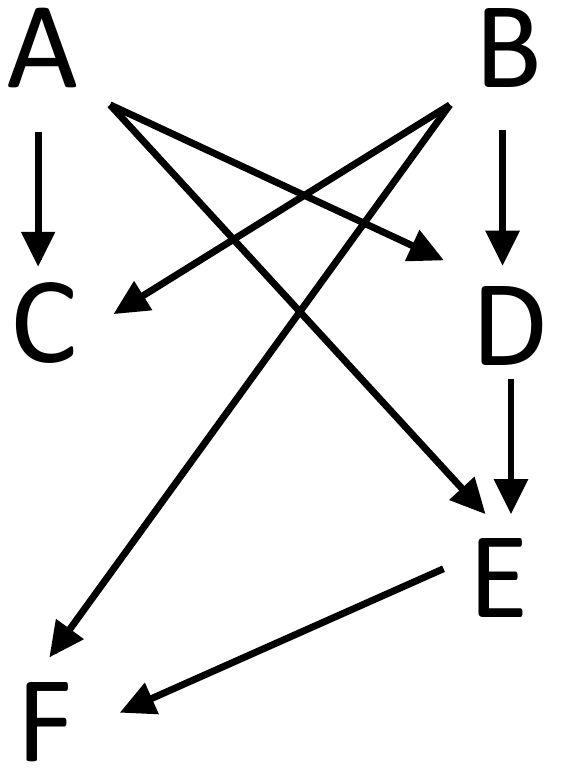}
\end{center}
\caption{A pedigree taken from  Kempthorne (\citeyear[p. 76]{kempthorne1969introduction}). The \emph{COP}, considering the nodes of the pedigree being either individuals or populations is given in the Appendix A3.\label{fig6}}
\end{figure}

When a pedigree contains the relationship between strains, the calculation of $f^*_{XY}$, the expected COP between two random individuals taken from each of two strains is simpler to calculate than $f_{XY}$ because inbreeding of a common ancestors takes no part, only allele frequencies. If the frequency of a particular allele \textit{A}  in two parent strains \textit{X}  and \textit{Y}  are $p_X$ and $p_Y$ respectively, then, assuming random mating, no mutation and no random drift, the frequency of that allele in the $F_1$ strain \textit{W} is 

\begin{equation}
p_W=(p_X+p_Y)/2
\label{main}
\end{equation}
thus, starting from a population of unrelated terminal ancestors with two alleles each at equal frequencies at a particular locus, it is relatively simple to track down the allele contribution of each terminal ancestor to any descendant using (\ref{main}). In the pedigree in Fig.~\ref{fig6} there are only two ancestors and we first build the allele composition of each terminal ancestor as:

\begin{eqnarray*}
X=[1/2 \  \  \ 0 ]  \\
Y=[0 \ \  \ 1/2 ],
\label{vec}
\end{eqnarray*}
where the $1/2$ comes from the fact that we will track temporarily only one allele from each terminal ancestor. From these, the allele frequencies of the selected alleles in the rest of strains in the pedigree can be calculated using (\ref{main}):
\begin{eqnarray*}
 W=(X+Y)/2 = [1/4 \ \ 1/4]\\
 Z=(X+W)/2 = [3/8 \ \ 1/8],
\end{eqnarray*}
and the probability of coincidence at the chosen allele from \textit{X} or the chosen allele from \textit{Y} is the cross product of their composition vectors, that is, $W' Z = 1/8$. Finally, the COP between strains \textit{W} and \textit{Z} is twice this product since terminal ancestors were heterozygous:
\begin{equation*}
f_{WZ}= 2W'Z=\frac{1}{4}
\end{equation*}

If terminal ancestors are homozygous and isogenic, then the vectors of terminal ancestors must be defined as:

\begin{eqnarray*}
X=[1 \  \  \ 0 ] \\
Y=[0 \ \  \ 1 ],
\end{eqnarray*}
and the expression for COP would be just $f_{WZ}=W'Z$ in this case. A matrix- based algorithm to calculate the COP between strains is proposed in Appendix A2 and a complete example is provided in Appendix A3.

\section{Discussion}

The calculation of the COP is based in calculating the probability that an ancestor has inherited a particular allele to two individuals. If alleles are assumed to descend through the nodes of a pedigree in a cascade, as it occurs when the nodes of a pedigree are single individuals, then the traditional formulae applies. Nevertheless, if nodes are strains, this affects how alleles descend through the pedigree, which is specially true if we add selection pressure at some or all nodes. The COP between all $21$ possible pairs in pedigree in Fig.~\ref{fig6} showed that a third of them yielded different result by assuming nodes are individuals or strains (see Appendix A3).

The idea behind the COP is built on expected values or frequencies, that can be estimated if the two individuals to compare are selected at random, thus, it seems unharmful to select (conceptually) one individual at random from each strain and build a genealogy among them that is identical to the pedigree for the strains where they come.This assumption is perhaps the origin of the mistake committed in using methods for pedigreed individuals in plant breeding. Nevertheless, as we have shown, the COP between two strains is just the cross product of their relative frequencies, due to an independence that may not exist if they were two individuals. 

Compelling evidence that the use of formulae designed for animal breeders must be used with reserve in plant breeding comes from a property of the COP. The COP of an individual with itself is:
\begin{equation}
f_{XX}=\frac{1}{2}(1+F_X)
\label{disc}
\end{equation}
(see \cite{falconer1996introduction}, Eq. 5.4 and \cite{bernardo2002breeding}, (Eq. 2.11). This poses some problems in plant breeding, for instance, assume \textit{X} is a fully inbred strain with two alleles, thus it has the two genotypes $[AA,BB]$ at equal expected frequencies. According to Eq. (\ref{disc}), $F_X = 1$ and the coefficient of parentage of strain \textit{X} with itself would be $f^*_{XX} = (1/2)(1+1) = 1$. Nevertheless if strain \textit{X} is crossed with itself by open pollination there is a $1/2$ probability that two identical alleles met in the same locus, therefore $f^*_{XX} = 1/2$.   It is clear that in this example $f^*_{XX}$ and $f_{XX}$ will coincide if we define a cross of a strain with itself as selfing each of the individuals of the strain, but keeping the usual definition of cross (open pollination) when the two strains are different.  This has never been discussed to the level of attention it deserves.

\section{Conclusions}

We have provided mathematical proof and via some examples that the COP between strains requires different calculations. Pedigrees in plant breeding are complex, since many times they involve backcrosses that affect the level of inbreeding of strains. The traditional calculations in animal breeding when extended to plant breeding resulted then in cumbersome calculations. This paper shows that the COP in pedigreed strains should be a simple process, as long as the assumptions of the Hardy-Weinberg equilibrium are observed.

\section{Appendix}
\subsection{A1}

Here we show that the COP between two strains according to the traditional calculations will sometimes overestimate the true COP, that is, $f_{XY} \ge f^*_{XY}$. Let \textit{X} and \textit{Y} be two arbitrary individuals in some arbitrary pedigree. Let \textit{A} be the event ``\textit{X} has allele \textit{W}" and \textit{B} the event ``\textit{Y} has allele \textit{A}", with respective probabilities $P(A)$ and $P(B)$. Independently of the topology of the pedigree there are only two possibilities: $(i)$ either one of \textit{X} and \textit{Y} is the ancestor of the other or $(ii)$ any other situation in which none of them can transmit allele \textit{W} to the other. Without loss of generality, assume that if the case is $(i)$ then \textit{X} is ancestor of \textit{Y}. 

If the case is $(i)$, then the probability that \textit{Y} has allele \textit{W} increases if \textit{X} has it for obvious reasons, that is, $P(B | A) \ge P(B)$. If the case is $(ii)$, then the fact that \textit{X} has allele \textit{W} does not change the likelihood that \textit{Y} has it, thus in this case, $P(B | A) = P(B)$.  Putting the two results together yield:

\begin{equation*}
P(B | A) \ge P(B)
\label{unq0}
\end{equation*}
 Since

\begin{equation*}
P(B | A)=\frac{P(A B)}{P(A)}
\end{equation*}
where $P(AB)$ is the probability that both \textit{X} and \textit{Y} have allele \textit{W}, it is implied that 

\begin{equation*}
\frac{P(A B)}{P(A)} \ge P(B)
\end{equation*}
or

\begin{equation}
P(A B) \ge  P(A) P(B)
\label{unq}
\end{equation}
that is, if \textit{X} and \textit{Y} are both individuals, the probability that both  have allele \textit{W} is greater or equal than the product of their individual probabilities of having allele \textit{W}.

On the other hand, if \textit{X} and \textit{Y} are strains, the probability that we take a random individual from population \textit{X} and another from population \textit{Y} and both have allele w at a random locus, is just the product of the probabilities that each individual has allele \textit{W}, that is, $P(A) P(B)$. Therefore, the left side of (\ref{unq}) is the probability that two individuals have allele \textit{W} that is, $f_{XY}$ whereas the right side is the probability that two randomly selected individuals from two strains have allele w which is $f^*_{XY}$. It follows directly that $f_{XY} \ge f^*_{XY}$.

\subsection{A2}

This is a simple algorithm to calculate the COP between strains in the absence of selection.

Assume we have a pedigree involving $N$ strains, labelled $1,2,3,\ldots, N$ Construct a matrix $\mathbf{P}$ with elements $p_{ij}$ such:

\begin{eqnarray*}
p_{ij}&=&1 \ \ \ \text{if strain $i$ is parent of $j$}\\
&=& 0 \ \ \ \text{otherwise}
\end{eqnarray*}
This will yield an $N$ by $N$ matrix whose columns add up to $0$ for the terminal ancestors and $2$ for those strains that are descendants of the terminal ancestors.
The matrix
$$
\mathbf{Q}=(\mathbf{I}-(1/2)\mathbf{P})^{-1}
$$
contains elements $q_{ij}$ such that $q_{ij}$ is the contribution of alleles of strain $i$ to strain $j$. In this matrix we must eliminate the contribution of those strains that are not terminal ancestors, that is, we need $q_{ij}=0$ if $i$ is not one of the terminal ancestors. To achieve this, we need to build a matrix $\mathbf{E}$ with elements $e_{ij}$ such that

\begin{eqnarray*}
e_{ii}&=&1  \ \ \ \text{if strain $i$ is a terminal ancestor}\\
&=& 0 \ \ \ \text{otherwise}
\end{eqnarray*}
that is, $\mathbf{E}$ is a diagonal matrix with a $1$ in position $e_{ii}$ if $i$ is the index of one of the terminal ancestors. The product $\mathbf{EQ}$ is a matrix where column $j$ contains twice the contribution of a particular allele of every terminal parent to that strain $j$ . The COP* is then

$$
COP^*=\frac{1}{2} (\mathbf{EQ})'(\mathbf{EQ})
$$
since $\mathbf{E'E}=\mathbf{E}$ we finally arrive to:
$$
COP^*=\frac{1}{2} \mathbf{Q'EQ}
$$
or just $\mathbf{Q'EQ}$ if terminal ancestors are homozygous. We develop a full exercise in appendix C.

\subsection{A3}

Complete example. This example is based on Fig.~\ref{fig6} taken from \cite[p. 76]{kempthorne1969introduction} Terminal ancestors are strains \textit{A} and \textit{B} that are assumed to be heterozygous and unrelated. 

Matrix $\mathbf{P}$ is equal to:

     $$\bordermatrix{\text{}&\text{A}&\text{B}&\text{C}&\text{D} &\text{E} &\text{F} \cr
                &  0 & 0 & 1 & 1 & 1 & 0  \cr
                &  0 & 0 & 1 & 1 & 0 & 1  \cr
        \mathbf{P} =        &  0 & 0 & 0 & 0 & 0 & 0  \cr
                &  0 & 0 & 0 & 0 & 1 & 0  \cr
               &  0 & 0 & 0 & 0 & 0 & 1  \cr
               &  0 & 0 & 0 & 0 & 0 & 0  \cr}
               $$
          
    Matrix $\mathbf{Q}=(\mathbf{I}-\frac{1}{2} \mathbf{P})^{-1}$: 
     
     $$\bordermatrix{\text{}&\text{A}&\text{B}&\text{C}&\text{D} &\text{E} &\text{F} \cr
                &  1 & 0 & 1/2 & 1/2 & 3/4 & 3/8  \cr
                &  0 & 1 & 1/2 & 1/2 & 1/4 & 5/8  \cr
        \mathbf{Q} =        &  0 & 0 & 1 & 0 & 0 & 0  \cr
                &  0 & 0 & 0 & 1 & 1/2 & 1/4 \cr
               &  0 & 0 & 0 & 0 & 1 & 1/2  \cr
               &  0 & 0 & 0 & 0 & 0 & 1 \cr}
               $$
Matrix $\mathbf{E}$:

 $$\bordermatrix{\text{}&\text{A}&\text{B}&\text{C}&\text{D} &\text{E} &\text{F} \cr
                &  1 & 0 & 0 & 0 & 0 & 0  \cr
                &  0 & 1 & 0 & 0 & 0 & 0  \cr
        \mathbf{E} =        &  0 & 0 & 0 & 0 & 0 & 0  \cr
                &  0 & 0 & 0 & 0 & 0 & 0  \cr
               &  0 & 0 & 0 & 0 & 0 & 0  \cr
               &  0 & 0 & 0 & 0 & 0 & 0  \cr}
               $$
               
               Matrix $\mathbf{EQ}$ is: 
     
     $$\bordermatrix{\text{}&\text{A}&\text{B}&\text{C}&\text{D} &\text{E} &\text{F} \cr
                &  1 & 0 & 1/2 & 1/2 & 3/4 & 3/8  \cr
                &  0 & 1 & 1/2 & 1/2 & 1/4 & 5/8  \cr
        \mathbf{EQ} =        &  0 & 0 & 0 & 0 & 0 & 0  \cr
                &  0 & 0 & 0 & 0 & 0 & 0 \cr
               &  0 & 0 & 0 & 0 & 0 & 0  \cr
               &  0 & 0 & 0 & 0 & 0 & 0 \cr}
               $$

Finally the matrix with the COP's is:

     $$\bordermatrix{\text{}&\text{A}&\text{B}&\text{C}&\text{D} &\text{E} &\text{F} \cr
                &  1/2 & 0 & 1/4 & 1/4 & 3/8 & 3/16  \cr
                &   & 1/2 & 1/4 & 1/4 & 1/8 & 5/16  \cr
        f^*_{XY} =        &  &  & 1/4 & 1/4 & 1/4 & 1/4  \cr
                &   &  &  & 1/4 & 1/4 & 1/4  \cr
               &   &  &  &  & 5/16 & 7/32  \cr
               &   &  &  &  &  & 17/64  \cr}
               $$

Meanwhile, the traditional \textit{COP} for individuals is:

     $$\bordermatrix{\text{}&\text{A}&\text{B}&\text{C}&\text{D} &\text{E} &\text{F} \cr
                &  1/2 & 0 & 1/4 & 1/4 & 3/8 & 3/16  \cr
                &   & 1/2 & 1/4 & 1/4 & 1/8 & 5/16  \cr
        f_{XY} =        &  &  & 1/2^* & 1/4 & 1/4 & 1/4  \cr
                &   &  &  & 1/2^* & 3/8^* & 5/16^*  \cr
               &   &  &  &  & 5/8^* & 3/8^*  \cr
               &   &  &  &  &  & 9/16^*  \cr}
               $$
where an `*' points out places where $f_{XY}$ and $f^*_{XY}$ yield different result. Observe that in all cases $f_{XY} \ge f^*_{XY}$.

\end{document}